\documentclass[11pt,american]{article}
\usepackage[latin9]{inputenc}
\usepackage{array}
\usepackage{varwidth}
\usepackage{amsmath}
\usepackage{amssymb}
\usepackage[a4paper]{geometry}
\usepackage{setspace}
\makeatletter

\DeclareFontEncoding{LGR}{}{}
\DeclareRobustCommand{\greektext}{%
  \fontencoding{LGR}\selectfont\def\encodingdefault{LGR}}
\DeclareRobustCommand{\textgreek}[1]{\leavevmode{\greektext #1}}

\providecommand{\tabularnewline}{\\}
\newenvironment{cellvarwidth}[1][t]
    {\begin{varwidth}[#1]{\linewidth}}
    {\@finalstrut\@arstrutbox\end{varwidth}}

\numberwithin{equation}{section}
\numberwithin{figure}{section}
\newenvironment{lyxlist}[1]
	{\begin{list}{}
		{\settowidth{\labelwidth}{#1}
		 \setlength{\leftmargin}{\labelwidth}
		 \addtolength{\leftmargin}{\labelsep}
		 }}
	{\end{list}}

\usepackage{paralist}
 
\usepackage{amsmath,amsthm,amssymb,mathtools,verbatim,url,physics}

\input{tcilatex}
\usepackage{titlesec}
\renewcommand\thesection{\arabic{section}.}
\renewcommand\thesubsection{\thesection\arabic{subsection}.}
\usepackage{textgreek}
\usepackage{pifont}
\newcommand{\xmark}{\ding{55}}%
\titleformat{\section}
  {\normalfont\bfseries\centering}{\thesection}{0.3em}{}
\titleformat{\subsection} {\normalfont\bfseries}{\thesubsection}{0.3em}{}
\usepackage{chngcntr}
\counterwithout{figure}{section}

\makeatletter
\renewcommand{\theequation}{\arabic{equation}}
\makeatother

\usepackage{tikz}
\usetikzlibrary{arrows.meta}

\makeatother

\usepackage{babel}
\begin{document}
\title{\textbf{A Heptalemma for Quantum Mechanics}\vspace{-0.1cm}
}
\author{{\Large John B. DeBrota \& Christian List\thanks{This version: 11 March 2026. Address for correspondence: Munich Center
for Mathematical Philosophy, LMU Munich. Both authors contributed
equally to this work. We thank Chris Fuchs, Stephan Hartmann, Levin
Hornischer, Beatrice Nettuno, Hrvoje Nikoli\'{c}, Daniele Oriti, Huw
Price, Simon Saunders, Blake Stacey, Ken Wharton, and the participants
of a research seminar at the MCMP in November 2025 for helpful conversations
and feedback.}}}
\date{{\small\vspace{-1cm}
}}

\maketitle
\noindent\noindent\rule{\linewidth}{0.4pt}

\noindent We present a seven-pronged no-go result for quantum mechanics:
a ``heptalemma''. It shows that seven initially plausible theses
about physical reality are jointly inconsistent \linebreak{}
with the predictions of quantum mechanics, while any six are jointly
consistent. We must then decide which theses to retain and which to
give up. Since different interpretations of quantum mechanics entail
different responses to the heptalemma, we get a novel taxonomy of
such interpretations. Beyond the application to quantum mechanics,
the heptalemma offers a general diagnostic criterion for determining
whether a given scientific domain should count as classical or not,
and if not, how it departs from classicality.

\noindent\noindent\rule{\linewidth}{0.4pt}

\section{Introduction}

It is widely appreciated that quantum mechanics gives us a picture
of physical reality that goes against some key tenets of a classical
physical worldview. Over the years, a series of paradoxes and no-go
results, including the Einstein-Podolsky-Rosen (EPR) paradox (Einstein
\emph{et al.} 1935), Bell's theorem (Bell 1964, 1985), the Kochen-Specker
theorem (Kochen and Specker 1967, Budroni \emph{et al.} 2022), and
the thought experiment of ``Wigner's friend'' (Wigner 1961, Schmid
\emph{et al.} 2023), have helped physicists to identify the ways in
which quantum-mechanical predictions are in tension with various traditional
or intuitive assumptions about the nature of the physical world.\footnote{Indeed, deriving no-go theorems for quantum mechanics is, by now,
a broader tradition. For a comprehensive review, see Leifer (2014).} But although those tensions are well known, there is still no consensus
on how to respond to them. What picture of reality best fits quantum
mechanics?

In this paper, we will present a new no-go result that we hope will
shed further light on this question. We will state seven initially
plausible theses that may all be viewed as key ingredients of a pre-quantum-mechanical
worldview \textendash{} the sort of worldview Einstein might have
subscribed to \textendash{} but which are jointly inconsistent with
the predictions of quantum mechanics. Since any six of the seven theses
are jointly consistent, our result admits seven distinct ``escape
routes'' (strictly speaking, families of escape routes), depending
on which thesis we reject. The result therefore forces us to decide
which theses to retain and which not. We call this a ``heptalemma
for quantum mechanics''. Different interpretations of quantum mechanics
can be shown to correspond to different ``horns'' of this heptalemma,
and so we get a novel taxonomy of interpretations of quantum mechanics.

Formally, the heptalemma can be viewed as an extension of Bell's theorem,
obtained via fine-graining some of Bell's assumptions. It therefore
stands in the tradition of earlier such refinements in the literature
(notably Jarrett 1984, where one of Bell's assumptions, locality,
is split into two assumptions, parameter independence and outcome
independence). We do something similar with Bell's assumption of realism
(sometimes also known as outcome definiteness or counterfactual definiteness),
but our philosophical ambition goes further. The goal is to offer
a detailed map of different ways in which the quantum-mechanical picture
of reality departs from the classical one.

The paper is structured as follows. In Section 2, we briefly review
Bell's theorem as a background to our analysis. In Section 3, we state
the seven theses and show how they lead to a heptalemma. In Section
4, we discuss the seven logically possible escape routes, and in Section
5, we say more about where three particularly challenging interpretations
of quantum mechanics \textendash{} Copenhagen, Everett, and QBism
\textendash{} fit into the resulting picture. In Section 6, we explain
how the heptalemma may be used as a criterion for determining which
scientific domains should count as classical and which not, and for
diagnosing different departures from classicality more broadly. In
Section 7, finally, we conclude.

\section{Bell's theorem}

Bell's theorem can be motivated by reference to the famous Einstein-Podolsky-Rosen
paradox (EPR) paradox (Einstein \emph{et al.} 1935). Recall that,
in quantum mechanics, it is impossible to measure an individual particle's
position and momentum simultaneously with perfect precision, by Heisenberg's
uncertainty principle. Several of the founders of quantum mechanics,
notably Bohr and Heisenberg, took this and other aspects of the new
theory to suggest that there is no fact of the matter about a particle's
position or momentum before or unless we perform a measurement. (Perhaps
the particle's position or momentum will become ``real'' only in
the act of measurement or is only a meaningful quantity with respect
to the particular experimental arrangement necessary to elicit a value.)
Einstein, Podolsky, and Rosen presented a thought experiment that
challenges this ``no fact'' assumption. Suppose we send two maximally
entangled particles in opposite directions and let each of them travel
very far, in principle even for light-years. The entanglement of the
two particles implies that if an observer were to measure the position
of one of them, they would be able to predict the outcome of a position
measurement of the other with perfect precision. Similarly, if an
observer were to measure the momentum of one of the particles, they
would be able to predict the outcome of a momentum measurement of
the other equally precisely. On the assumption that physics admits
no ``spooky action at a distance'', especially no transmission of
information faster than the speed of light, there is no way any measurement
performed on the first particle could instantaneously affect the second
particle. It should make no difference to the second particle whether
the observer measures the position or the momentum of the first particle.
Einstein, Podolsky, and Rosen concluded that the perfect predictability
of the measurement outcome for the second particle can only be explained
if there are some pre-existing facts about what those measurement
outcomes would be.\footnote{Here they invoke the principle that ``{[}i{]}f, without in any way
disturbing a system, we can predict with certainty (i.e., with probability
equal to unity) the value of a physical quantity, then there exists
an element of physical reality corresponding to this physical quantity''
(Einstein \emph{et al.} 1935, p. 777).} This goes against Bohr's and Heisenberg's ``no fact'' assumption.
Einstein, Podolsky, and Rosen took this to suggest that quantum mechanics
is incomplete. It seems there must be some hidden variables by which
the measurement outcomes have been determined in advance.\footnote{The EPR paradox is often misrepresented. For an exceptionally clear
exposition, we recommend Stacey (2018).}

This is where Bell's theorem comes into the picture. It shows that
the assumption of hidden variables would not resolve the EPR paradox
if we retain the other assumptions of the thought experiment. Rather,
we face a general no-go result. Suppose we make the following assumptions
about a physical system:

\medskip{}

\noindent\textbf{Realism:} Any observer's measurement outcomes correspond
to objective facts about the system over which probabilities are well-defined.

\medskip{}

\noindent\textbf{Locality:} Measurements performed in one place do
not instantaneously bring about effects in a distant place.

\medskip{}

\noindent\textbf{Measurement independence:} Measurement choices (e.g.,
the settings of a measurement apparatus) are probabilistically independent
of each other and of any states of the system that are being measured.
(Informally, observers can freely choose their measurement settings,
independently of one another and of the system in question.)

\medskip{}

\noindent Bell's theorem shows that these three assumptions are jointly
inconsistent with the predictions of quantum mechanics. More precisely,
Bell's theorem establishes the following:
\begin{itemize}
\item If a system meets these three assumptions, then the system will satisfy
certain constraints on the correlations between the outcomes of different
measurements that we could perform on that system \textendash{} the
so-called Bell inequalities; and
\item quantum mechanics predicts violations of those inequalities.
\end{itemize}
One can think of the Bell inequalities as expressing upper bounds
on the correlations that can occur in a system satisfying the three
assumptions. Slightly more technically, each of the terms of a particular
Bell inequality known as the CHSH (Clauser-Horne-Shimony-Holt) inequality
corresponds to a different pair of measurements that could be performed
on the given system (Clauser \emph{et al.} 1969).\footnote{The satisfaction of the CHSH inequality, in turn, is equivalent to
the existence of a single joint probability distribution for all measurement
outcomes (Fine 1982).} Under Bell's assumptions, the expectations of four distinct such
measurement pairs combined in a particular way are bounded above by
a value that quantum-mechanical predictions can violate. And so it
follows that quantum-mechanical systems cannot satisfy all of realism,
locality, and measurement independence. At least one of these assumptions
must be false.

To illustrate this result, let us represent something like the EPR
setting with the help of a simple model. Suppose that two particles
are sent in opposite directions (``left'' and ``right''), starting
from a common source, and suppose that one observer, Alice, performs
a measurement on the left particle, while another observer, Bob, performs
a measurement on the right particle. The setup includes five variables,
$\lambda,A,B,a,b$, where $\lambda$ stands for the state of the common
source, $a$ and $b$ stand for Alice's and Bob's measurement settings
(broadly analogous to the choice between measuring position or measuring
momentum), and $A$ and $B$ stand for Alice's and Bob's measurement
outcomes. Bell's assumptions entail that the probabilistic dependencies
between the five variables are as shown by the directed acyclic graph
in Figure 1, where arrows represent probabilistic dependencies between
variables.
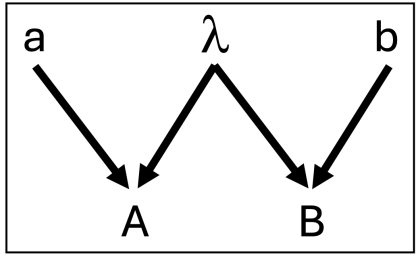
\begin{figure}[h]
\begin{centering}
\caption{Bell's assumptions illustrated}
\par\end{centering}
\centering
\begin{tikzpicture}[>=Triangle, thick, every node/.style={font=\Large}]
\draw[line width=0.8pt] (-0.5,-0.5) rectangle (4.9,2.1);
\node (a) at (0,1.7) {$a$};
\node (lambda) at (2.3,1.7) {$\lambda$};
\node (b) at (4.5,1.7) {$b$};
\node (A) at (1.1,0) {$A$};
\node (B) at (3.4,0) {$B$};
\draw[->, line width=1.5pt] (a) -- (A);
\draw[->, line width=1.5pt] (lambda) -- (A);
\draw[->, line width=1.5pt] (lambda) -- (B);
\draw[->, line width=1.5pt] (b) -- (B);
\end{tikzpicture}
\end{figure}
 In particular, because of locality, Alice's measurement outcome,
$A$, depends on the state of the common source ($\lambda$) and her
chosen measurement setting ($a$), but not on anything that happens
on Bob's side (such as $b$ or $B$). Similarly, Bob's measurement
outcome, $B$, depends on the state of the common source ($\lambda$)
and his measurement setting ($b$), but not on anything that happens
on Alice's side (such as \emph{a} or \emph{A}). Finally, because of
measurement independence, Alice's and Bob's measurement settings,
$a$ and $b$, are independent from each other and from the common
source of the particles ($\lambda$).\footnote{Readers might wonder how this model invokes Bell's assumption of realism.
That assumption allows us to postulate that there are objective facts
about the five variables and their probabilistic dependencies.} This dependency structure can be shown to satisfy the CHSH inequality,
as explained in detail in the appendix. Of course, there will be some
correlations in the present system, but in the simple model of Figure
1, these satisfy the relevant constraints. For example, there is a
correlation between $A$ and $B$, since they are both dependent on
$\lambda$. At the same time, $A$ and $B$ will be conditionally
independent, given their common causes.\footnote{We can think of the directed acyclic graph in Figure 1 as being associated
with a Bayesian network. Locality and measurement independence can
then be viewed as implications of the parental Markov condition applied
to the given network structure. Generally, in a Bayesian network satisfying
the parental Markov condition, any variable is independent of its
non-descendants, given its parents (Pearl 2000).} The crux is that quantum mechanics predicts breaches of the implied
constraints on the correlations, and so the given model is inconsistent
with quantum mechanics. Since the model accurately represents the
implications of Bell's assumptions, at least one of the three assumptions
must be false.

It is important to note that there is nothing incoherent about Bell's
assumptions \emph{by themselves}. In a classical mechanical system,
say a system of billiard balls that behave according to the laws of
classical physics, the assumptions can be unproblematically met. There
could be a classical physical system that is accurately represented
by the setup of Figure 1. What Bell's theorem establishes is that
the three assumptions are jointly inconsistent with quantum mechanics.
In effect, quantum mechanics implies that reality behaves non-classically.

Bell's theorem is useful not only for diagnosing how quantum mechanics
departs from the classical picture of reality, but also for generating
a taxonomy of different interpretations of quantum mechanics. These
can be classified in terms of which of Bell's assumptions they give
up. Let us briefly illustrate this by reference to a few examples.
One class of interpretations gives up the assumption of locality.
The de Broglie-Bohm interpretation (Goldstein 2025), for instance,
takes this route, by postulating that quantum-mechanical systems behave
deterministically based on the \emph{global} configuration of underlying
particle positions. Giving up locality, however, amounts to admitting
what Einstein called ``spooky action at a distance''. A second class
of interpretations gives up measurement independence. So-called superdeterministic
interpretations fall into this class. They postulate that there are
dependencies between measurement choices and the states of the system
that are being measured: different observers' measurements are correlated
due to underlying initial conditions. This kind of interpretation
is sometimes characterized as denying that observers have free will
with respect to the choice of measurement settings \textendash{} an
implication that some scholars see as a feature and others as a bug.
There is also a different way of relaxing measurement independence,
which seeks to preserve the observers' freedom by postulating retrocausality
(see, e.g., Price and Wharton 2015). A third class of interpretations,
finally, gives up realism as it is defined in Bell's theorem. This
is exemplified by the kind of view associated with Bohr and Heisenberg,
often called the Copenhagen interpretation, though there are different
variants of it. As already noted, Bohr and Heisenberg thought that
there is no fact of the matter about certain properties of a physical
system until these are explicitly measured. However, a very diverse
range of theories is usually grouped together under the umbrella of
the denial of Bell's realism assumption, and as we will see, there
are many distinct ways in which one might relax that assumption. Accordingly,
we will show that one can arrive at a more fine-grained taxonomy.\footnote{Importantly, of course, the literature already contains several more
nuanced no-go results in the vicinity of Bell's theorem. For example,
in a similar spirit to Bell's theorem, Greenberger \emph{et al. }(1989)
derive a deterministic contradiction between local realism and quantum
predictions for three or more particles; Hardy's paradox (Hardy 1992)
shows a contradiction between local realism and quantum predictions
using a single set of events; and Leggett's theorem (Leggett 2003)
cuts locality and realism along different seams, ruling out a different
class of hidden-variable models if a stronger realism condition is
adopted. Subject to a preparation independence assumption, the Pusey-Barrett-Rudolph
theorem (Pusey \emph{et al.} 2012) rules out \emph{purely} epistemic
interpretations of the wavefunction. The Colbeck-Renner theorem (Colbeck
and Renner 2011) shows, under assumptions of measurement independence
and no-signaling, that no hidden variable extension of quantum theory
can improve outcome predictions.}

\section{A heptalemma}

We will show that quantum mechanics faces a seven-pronged no-go result
that can be viewed as a refinement of Bell's theorem. Specifically,
there are seven initially plausible theses about physical reality
which are jointly inconsistent with the predictions of quantum mechanics
and of which any six are consistent with those predictions.\footnote{While some of our theses are inherited from Bell's theorem, others
are variants of related theses from the philosophical literature,
especially in two philosophical no-go results in List (2025) and Fine
(2005), respectively. We will say more about the connections with
that literature below.}

Our first step is to unpack the formulation of realism in Bell's theorem.
We propose that this actually expresses the conjunction of four theses,
which are jointly sufficient for realism in the original sense used
in Bell's theorem. We will first state the four theses and then make
some further explanatory comments.

\medskip{}

\noindent\textbf{Measurement realism:} Any observer's measurement
outcomes correspond to facts over which there are well-defined probabilities,
where there is no presumption whether the facts are objective or subjective,
absolute or relative, coherent or incoherent.

\medskip{}

\noindent\textbf{Non-relationalism:} Facts, including all to which
any observer's measurement outcomes correspond, are non-relational:
when they obtain, they obtain \emph{simpliciter}, not relative to
something else.

\medskip{}

\noindent\textbf{Non-fragmentation:} Any possible world is coherent,
i.e., the total collection of facts that hold at that world can be
jointly instantiated.

\medskip{}

\noindent\textbf{One world: }Reality is exhausted by one objective
world, which can be defined as the total collection of facts that
hold at that world.

\medskip{}

The first thesis \textendash{} measurement realism \textendash{} asserts
only a very undemanding form of realism. It says that any observer's
measurement outcomes correspond to \emph{some} aspects of reality
\textendash{} some ``facts'' \textendash{} while still leaving open
the precise nature of those facts. In particular, measurement realism
says nothing about whether measurement outcomes correspond to objective
or subjective facts, to absolute or relative ones, to jointly coherent
or incoherent ones. So, as far as measurement realism is concerned,
the postulated facts could be facts in an extraordinarily weak sense:
they could be subjective rather than objective, relative rather than
absolute, for instance relative to an observer or some other physical
system, and they could even be incoherent rather than coherent. The
next three theses then successively constrain the nature of the postulated
facts, demanding that they must be absolute (not relative), jointly
coherent (not possibly incoherent), and objective (not subjective).
By way of contrast, Bell's original assumption of realism already
presupposes all of these constraints.

In more detail, non-relationalism asserts that any fact is of the
``absolute'' form ``such and such is the case'', not of the ``relative''
form ``such and such is the case, relative to something else''.
As Wittgenstein noted in his \emph{Tractatus}, a fact is something
that is the case. Examples of facts are the following: water is H$_{2}$O;
a hydrogen atom contains one proton; $2+2=4$; humans live on Earth.
By accepting these as facts, we take them to hold \emph{simpliciter},
not relative to something else. For example, humans live on Earth,
full stop, not just relative to something else. If facts were relational,
facts could only be properly specified as ordered pairs of the form
$\left\langle A,B\right\rangle $, where $A$ is something that is
the case relative to $B$. Thus facts would always be \emph{dyadic}.
On the standard view according to which facts are non-relational,
by contrast, any fact is \emph{monadic}: a fact is something ($A$)
that is the case \emph{simpliciter}; no relativization to anything
else (any $B$) is needed.

Non-fragmentation then states that any possible world \textendash{}
i.e., any way a world could be \textendash{} is coherent. We should
think of any world \textendash{} whether actual or possible \textendash{}
as a coherent and complete collection of facts. Equivalently, the
total collection of facts that obtain at any given world can be coherently
coinstantiated. This also goes back to one of Wittgenstein's famous
ideas from his \emph{Tractatus}, namely the idea that ``{[}t{]}he
world is everything that is the case'', which he further clarifies
by saying ``{[}t{]}he world is the totality of facts'' and ``the
totality of facts determines both what is the case, and also all that
is not the case'' (Wittgenstein 1922, 1, 1.1, 1.12). The background
assumption is, of course, that the totality of facts making up the
world is coherent. Otherwise we would not be dealing with a \emph{possible}
world. The assumption of non-fragmentation is so widely accepted that
it is seldom even explicitly acknowledged. After all, the idea of
coherent worlds is central to science, logic, and philosophy.

The thesis of one world, finally, asserts a central tenet of an objectivist
worldview: reality is exhausted by one objective world, not multiple
``subjective'' worlds. It is not the case that different observers
occupy their own distinct subjective worlds. This assumption is also
almost unanimously accepted throughout science and philosophy. As
we will see, even the Everett interpretation of quantum mechanics,
which is sometimes (perhaps misleadingly) described as a ``many worlds''
interpretation, can be viewed as retaining the assumption of one world.

What Bell's theorem shows, in effect, is that any system satisfying
these four theses together with locality and measurement independence
must satisfy Bell inequalities. And since quantum-mechanical systems
can violate Bell inequalities, we can infer that the conjunction of
the four theses conflicts with locality and measurement independence.
Strictly speaking, however, the conflict arises only in conjunction
with a further thesis, which is usually presupposed tacitly:\medskip{}

\noindent\textbf{Non-solipsism:} Reality admits more than one observer.

\medskip{}

If there were only one observer, we would not even be able to construct
the kinds of scenarios in which different observers could each choose
their measurement settings independently and in which the issue of
Bell-inequality-violating correlations could arise.

Combining the foregoing theses we get the following no-go result:\medskip{}

\noindent\textbf{Heptalemma for quantum mechanics: }Given the predictions
of quantum mechanics, the following seven theses are mutually inconsistent:
measurement realism, non-relationalism, non-fragmentation, one world,
locality, measurement independence, and non-solipsism. Any subset
of these theses is mutually consistent.

\medskip{}

As already noted, the easiest way to derive this result is via the
observation that four of our theses jointly imply realism in Bell's
original sense, and that Bell's original theorem presupposes non-solipsism.
When we further add locality and measurement independence, the no-go
result then follows from Bell's theorem. Like Bell\textquoteright s
original theorem, the present result establishes that the relevant
theses are inconsistent \emph{with the predictions of quantum mechanics}.
The seven theses \emph{by themselves} are mutually consistent. We
will return to the significance of this point in Section 6.

Interestingly, our no-go result can also be seen as a quantum-mechanical
analogue of some recent no-go results from the philosophical literature
on first-personal facts. Examples of first-personal facts are ``I
am having such-and-such experiences'' or ``I occupy such-and-such
position in the world''. First-personal facts are non-objective insofar
as they are non-invariant under changes in the subjective perspective
from which they hold. Whether it is true that I am having such-and-such
experiences obviously depends on who I am and which perspective I
occupy. There are at least two no-go results concerning first-personal
facts in the recent philosophical literature. One, called a ``quadrilemma
for theories of consciousness'' (List 2025), asserts that realism
about first-personal facts is jointly inconsistent with non-solipsism,
non-fragmentation, and one world, where these theses are essentially
the same as in the present heptalemma. This ``quadrilemma'' presupposes
that first-personal facts are non-relational, and so it implicitly
relies on the thesis of non-relationalism as well.\footnote{According to realism about first-personal facts as discussed in List
(2025), it is not merely the case that ``relative to Christian's
perspective, I am having Christian's experiences'', but ``I am having
Christian's experiences'' \emph{simpliciter}.} Another related no-go result, due to Fine (2005), asserts a conflict
between realism about first-personal facts and three other theses
called ``first-personal neutrality'' (which broadly plays the role
of non-solipsism), ``absolutism'' (broadly equivalent to non-relationalism),
and ``coherence'' (broadly equivalent to non-fragmentation). Fine's
result can be interpreted either as presupposing one world or as treating
the combination of non-fragmentation and one world as a single thesis.

The analogy between our heptalemma and those philosophical no-go results
goes as follows. Suppose we accept locality and measurement independence
in Bell's theorem. Then, given the predictions of quantum mechanics,
we must reject Bell's original assumption of realism. Suppose, however,
we would still like to retain measurement realism in the very undemanding
sense we have introduced, i.e., we would still like to postulate some
``facts'' \textendash{} however minimalistically understood \textendash{}
that correspond to each observer's measurement outcomes. And suppose
further that we do not want to go along the solipsistic route of assuming
that reality admits only a single observer. Then something about the
postulated facts must be non-classical, i.e., distinct from how Bell
or Einstein, Podolsky, and Rosen thought about those facts. Specifically,
the postulated facts must either be of a relational sort, or they
must fail to generate a coherent totality, or the total collection
of facts must be able to vary from observer to observer. In the first
of these cases, we have a breach of non-relationalism; in the second,
a breach of non-fragmentation; and in the last, a breach of one world.
This structurally resembles the above-mentioned no-go results for
first-personal facts. To see that any six of our seven theses are
jointly consistent with the predictions of quantum mechanics, we now
turn to the possible escape routes from the inconsistency.

\section{Escape routes from the heptalemma}

While the seven theses of the heptalemma are jointly inconsistent
with the predictions of quantum mechanics, any six can be jointly
satisfied. In other words, there are seven fundamentally distinct
ways to ``escape'' the heptalemma, depending on which thesis one
rejects. These seven escape routes correspond to different interpretations
of quantum mechanics, or more precisely different \emph{families}
of interpretations. Which theses an interpretation rejects and which
it salvages tells us something about its priorities in the negotiation
of an updated scientific worldview. In this section we discuss each
of the possible escape routes.

\subsection{Rejecting locality}

As we have seen, one reaction to Bell's theorem is to reject locality,
i.e., to allow measurements performed in one place to\emph{ }instantaneously
bring about effects in a distant place or to stand in correlations
that cannot be accounted for by purely local mechanisms. This, of
course, remains a consistent escape route from the heptalemma as well.

The idea that reality admits only local influences is deeply ingrained
in modern physics and philosophy. Physics, however, did not consistently
embrace the idea of locality until the late 19th century, and only
with Einstein did strict relativistic locality become foundational.
For example, Newton's principle of universal gravitation violates
locality, allowing for ``action at a distance''. But even then,
the intuition of locality was pervasive; Newton himself was dissatisfied
with the non-locality in his theory (Berkovitz 2007). In a letter
to Richard Bentley, he wrote:
\begin{quote}
``That gravity should be innate{[},{]} inherent and {[}essential{]}
to matter so that one body may act upon another at a distance through
a vacuum without the mediation of any thing else by and through which
their action or force {[}may{]} be conveyed from one to another is
to me so great an absurdity that I believe no man who has in philosophical
matters any competent faculty of thinking can ever fall into it.''
(Newton 1692/3, also quoted in Berkovitz 2007)
\end{quote}
Arguably, part of Einstein's triumph with his relativity theories
lies in the physical codification of this intuition of locality. In
philosophy, similarly, locality is a widely held assumption. In Hall's
influential analysis of the concept of causation as production, locality
is treated as a key defining condition of causal production (Hall
2004).

A relaxation of locality therefore comes with substantial metaphysical
cost. In its most clearcut form, it means admitting what Einstein
called ``spooky action at a distance''. While he did not live to
see Bell's theorem, it seems likely that Einstein would have entertained
almost any other option. Indeed, he argued that rejecting locality
imperiled the entire enterprise of science, writing that ``{[}t{]}he
complete suspension of this basic principle would make impossible
the idea of the existence of (quasi-) closed systems and, thereby,
the establishment of empirically testable laws in the sense familiar
to us\textquotedblright{} (Einstein 1948, p. 322, quoted and translated
in Howard 1985, p. 188). While many do not share these instincts,
locality is certainly not abandoned lightly.

Moreover, any rejection of locality must be done with care as quantum
mechanics does not permit \emph{signaling}, the transmission of information
faster than the speed of light. In other words, \emph{even if} there
is non-locality in a physical system, this property cannot be leveraged
to enable superluminal communication. We would thus require a form
of non-locality that carefully constrains the permissible action at
a distance. The onus seems to be on proponents of non-locality to
explain why there is non-locality and yet no signalling.

That said, several interpretations of quantum mechanics do reject
locality. First and most familiar is the de Broglie-Bohm interpretation,
already mentioned in Section 2. It postulates particles with individual
velocities that instantaneously depend on the full configuration of
every other, no matter how distant, in a manner derived from the physical
wavefunction. Another family of interpretations that reject locality
are the so-called generalized collapse models, including those developed
by Ghirardi, Rimini, Weber and others (Ghirardi \emph{et al.} 1986).
These interpretations posit a stochastic element in the dynamical
evolution of the wavefunction, such that its non-local collapses,
including those upon measurement, may be given a more familiar dynamical
explanation. The so-called transactional interpretation (Cramer 1986),
meanwhile, explains quantum correlations in terms of waves that propagate
both forward and backwards in time. The wavefunction realism of Ney
(2021), in some ways more parsimoniously, proposes that wavefunctions
are physically real and physically collapse. While this collapse is
non-local from the perspective of our usual three-dimensional space,
wavefunctions are not taken to \emph{inhabit} this three-dimensional
space, but ultimately \emph{give rise} to it: the (approximate) locality
of our experience is supposed to be an emergent phenomenon. Finally,
an interpretation recently developed by Barandes (2025) takes quantum
mechanics to be a kind of ``indivisible'' stochastic process theory.
The idea here is that the predictions of quantum mechanics are consistent
with a non-Markovian stochastic process theory, where hidden variables
are non-local, but, of course, signaling is still prohibited. There
are certainly more non-local interpretations \textendash{} this list
is not an exhaustive catalog. The point is that the escape route of
rejecting locality is not only consistent, but popular as well.

\subsection{Rejecting measurement independence}

A second escape route that we already mentioned in relation to Bell's
theorem is to reject measurement independence. This would permit dependencies
between Alice's and Bob's measurement choices as well as between the
choices of either and the system being measured. However, if Alice's
or Bob's choices depend on the choices of the other or on the system
itself, the sense in which these can be said to be \emph{choices}
at all is called into question. This challenge is not easy to rebut,
since the rejection of measurement independence seems to allow for
the possibility that the system in question influences the supposed
choices of measurements made upon it. As we also mentioned above,
denial of measurement independence is sometimes associated with a
denial that experimenters possess free will as far as their choices
of measurement settings are concerned. It can of course be contested
whether a lack of measurement independence really implies a lack of
free will and, if so, whether this speaks against the present escape
route. Be that as it may, measurement independence may be rejected
in the heptalemma, leaving behind six mutually consistent theses.

Taken to its extreme, one could even reject measurement independence
\emph{deterministically}, such that the choices of Alice and Bob are
determined \emph{with certainty} by the physical state of the system.
It is perhaps unsurprising that there are physicists and philosophers
who prefer interpretations of this kind, known as superdeterministic
interpretations, as this intuition amounts to a supreme dedication
to physical determinism. A recent view of this type is the cellular
automaton interpretation of 't Hooft (2016). Here it is imagined that
physical reality is fundamentally discrete, local, and deterministic
and that this state of affairs is best understood by analogy with
models of computation like cellular automata. Everything that happens
in reality, including any proposed Bell scenarios and choices of the
agents involved, is determined by the initial state of the universe.
Another variant of superdeterminism is discussed by Hossenfelder and
Palmer (2020) and does not insist on a discrete, computational analogy.
For an approach that rejects measurement independence without postulating
determinism, see, for example, the statistical contextual interpretation
(Kupczynski 2025).

Superdeterministic approaches give up the idea that measurement settings
are exogenous variables whose values can be ``freely'' chosen by
the relevant observers. By contrast, there is also an approach that
relaxes measurement independence while still treating measurement
settings as exogenous. It is the retrocausal approach defended, for
instance, by Price and Wharton (2015); for a survey, see Friederich
and Evans (2023). This approach accounts for probabilistic dependencies
between the state of the common source and the observers\textquoteright{}
measurement settings by allowing the latter to influence the former,
rather than the other way around. In effect, this reverses the direction
of some of the arrows in the kind of causal graph illustrated in Figure
1 above. This blocks the derivation of the Bell inequalities without
abandoning the intuitive idea that the observers\textquoteright{}
measurement choices are in some sense free. The cost is the admission
of retrocausality.

\subsection{Rejecting non-solipsism}

Rejecting non-solipsism is the first of five novel escape routes that
the heptalemma allows us to discuss. This route supposes that reality
\emph{does not} admit more than one observer, i.e., there is \emph{at
most} one real observer. This is a consistent, albeit rather extreme,
way out because it cuts off the thought experiment before it can even
begin. If there is only one observer, we cannot actually have Alice
\emph{and} Bob.

While rejecting non-solipsism is a logically possible escape route,
it does not appear that anyone seriously entertains it as a reasonable
response to the challenges of quantum mechanics.\footnote{Convivial Solipsism (Zwirn 2000) and the solipsistic hidden variable
model of Nikoli\'{c} (2012) both postulate multiple observers and
are therefore not true solipsisms under our definition. The latter
can be converted into a true solipsism by simply supposing that only
one observer exists. As presented, however, objective reality is associated
to each conscious observer, and so Nikoli\'{c}'s model is actually
a rejection of the one-world postulate.} That said, we note that QBism (Fuchs and Schack 2013, Fuchs and Stacey
2018) has been accused of being a solipsistic view (e.g., Norsen 2014).
This accusation is derived from QBism's suggestion that the quantum
formalism is a ``single-user theory'', i.e., that when it is used,
it concerns the actions and experiences of a single agent, the user.
While this reading of the formalism only \emph{requires} one observer,
QBists are certainly not supposing that there actually is only one.
Rather, the quantum formalism, according to QBism, may be adopted
by \emph{any} observer to help her navigate the consequences of her
actions. A solipsist could interpret quantum mechanics in a QBist
way, but QBism itself does not entail solipsism at all. We say more
about QBism in Section 5.

\subsection{Rejecting measurement realism}

Rejecting measurement realism is the second novel escape route and
the first that can be viewed as a refinement of one of the existing
routes out of Bell's theorem. As Bell's realism combines the theses
of measurement realism, non-relationalism, non-fragmentation, and
one world, rejection of any of these four corresponds to rejecting
Bell's realism. However, specifically rejecting one of these finer-grained
theses enables us to draw more precise distinctions and to compare
different reasons why realism in Bell's original sense might not hold.

One readily sees that there are two ways to reject measurement realism.
The first is to claim that measurement outcomes do not correspond
to facts in \emph{any} sense. It is not that facts about measurement
outcomes are subjective rather than objective, dyadic rather than
monadic, or fail to be coherent in the standard sense. They are simply
not facts \emph{at all.} Measurement outcomes are not something that
is the case, no matter how loosely you interpret the condition.

The second way to reject measurement realism supposes that outcomes
\emph{are} facts in one or more of the relevant senses, but that there
are no well-defined probabilities over them. An absurd caricature
of this second option could even claim that measurement outcomes are
objective facts in one coherent world, yet the quantum formalism does
not give us anything that is interpretable as probabilities over those
facts. But rejecting measurement realism in this way would have to
be accompanied by some account of why quantum mechanics is correct
even if the numbers it spits out are not interpretable as probabilities
at all.

Both ways of rejecting measurement realism have precedent. Quantum
Darwinism (Zurek 2009) is a modern example of a rejection of the first
kind. The idea is to try to solve the quantum measurement problem
by absorbing the concept of measurement into quantum dynamics. Towards
this, it is argued that the phenomenon of decoherence allows one to
understand measurement outcomes as quasi-stable aspects of the physical
unitary evolution. Zurek writes: ``This transition from uncertainty
(initial presence of many branches \textendash{} potential for multiple
outcomes) to certainty (once a sufficiently long branch fragment becomes
known) accounts for perception of `collapse'{}'' (2009, p. 185).
Thus, while we have the perception of stable measurement outcomes,
there is no associated fact at the lowest level. We will argue in
Section 5.2 that the Everett interpretation (Wallace 2012, Vaidman
2021), also known as ``the many-worlds interpretation'' or ``the
relative state formulation'', is best understood, despite these names,
as rejecting measurement realism in the second way. Briefly, because
\emph{all} outcomes factually occur, there is no sense in which they
occur with well-defined probabilities. (In Section 5.2, we will also
consider some alternative ways of interpreting Everett.)

Although interpretationally quite far from Everett, one can make the
case that some variants of the Copenhagen interpretation reject measurement
realism too. However, as we elaborate in Section 5.1, Copenhagen is
not just one view; while all versions robustly reject Bell's realism,
it does not appear that all versions cleanly reject only one of the
four theses that constitute it. Finally, one might read into QBism
an inclination towards rejecting measurement realism, but as we argue
in Section 5.3, this is not what the conceptual thrust of QBism most
clearly points to.

\subsection{Rejecting non-relationalism}

If one wishes to preserve the relatively undemanding notion of realism
articulated by measurement realism, one can still deny one of the
three additional constraints that together entail Bell's original
realism. The first of these is non-relationalism. Rejecting non-relationalism
means admitting a nonstandard relational character for at least some
facts. This is a revision of the nature of a fact itself \textendash{}
it is not the case that when a fact obtains, it obtains \emph{simpliciter}.
Rejecting non-relationalism means that at least some facts are dyadic
in nature. On this view, a fact is not something that simply is the
case, but may be something that is the case only relative to something
else. Generically, any fact is of the form ``A is the case, relative
to B'', and so any collection of facts is a collection of ordered
pairs of the form $\left\langle A,B\right\rangle $. This move can
disable the derivation of Bell inequalities by massively enlarging
the sample space. Outcomes are no longer of the ``absolute'' or
``monadic'' form ``such and such outcome occurred'', but only
of the ``relational'' or ``dyadic'' form ``such and such outcome
occurred, relative to such and such relativization parameter'', where
the latter could be an observer or a system.

The best-known interpretation which takes this route is aptly named
Relational Quantum Mechanics (RQM) (Rovelli 1996, 2025). In RQM, the
state of a quantum system is inherently relational, i.e., the state
\emph{is} the relation between observer and system, or even more generally
between system and system. In this view, any physical system can have
a quantum state relative to any other \textendash{} so, there are
no stipulations on what could count as an ``observer'' \textendash{}
and there are no states in a non-relational sense. This commitment
to relationalism has been explicitly spelled out by Di Biagio and
Rovelli (2021). They write: ``Facts happen at every interaction,
but they are not absolute: they are relative to the systems involved
in the interaction'' (p. 29). The authors go on to develop an understanding
of what they call stable facts, which can be understood as facts whose
relativity can be mostly ignored. Although facts are intrinsically
dyadic, of the form $\langle A,B\rangle$, which can be read ``$A$
is the case relative to $B$'', a stable fact is one that is invariant
under changes in the second component; i.e., $\langle A,B\rangle$
is a stable fact if and only if it continues to hold whenever $B$
is replaced by any other admissible $B'$. RQM correctly identifies
itself as a realist view, moreover, explicitly adhering to the one-non-fragmented-world
picture when it describes itself as
\begin{quotation}
\noindent ``{[}r{]}ealist in the sense that it is not about agents,
beliefs, observers, or experiences: it is about real facts of the
world and relative probabilities of their occurrence. The ontology
is relational, in the sense that it is based on facts established
at interactions and are labelled by physical contexts.'' (Di Biagio
and Rovelli 2021, p. 8)
\end{quotation}
\noindent Thus RQM retains measurement realism, non-fragmentation,
and one world. There is also room to define variants of RQM with different
degrees of commitment to relationalism (see, e.g., Calosi and Riedel
2024, Stacey 2021).

Some other interpretations appear to reject non-relationalism too,
though they differ in what exactly they take outcomes or facts to
be relative to, for instance whether the relativization can be to
any system or merely to an observer. For example, Brukner (2018, p.
7) notes that a possible consequence of his no-go result is that ``there
cannot be facts of the world per se, but only relative to an observer.''
Healey's pragmatic interpretation (2012) takes a different route in
considering measurement outcomes to be non-relational and yet considering
quantum states to be agent-relative.

\subsection{Rejecting non-fragmentation}

Another way to retain measurement realism while giving up Bell's classical
realism is to reject non-fragmentation. Recall that non-fragmentation
asserts that any possible world is coherent in the sense that all
facts that obtain at it can indeed be jointly instantiated. Another
gloss on the same idea is that a world is a \emph{coherent totality.
}To reject this is to admit that while there might be locally coherent
subcollections of all the facts that hold at a given world, one cannot
demand that every fact is compatible with every other. This move allows
one to escape the heptalemma because it allows for the possibility
that in some circumstances, like the Bell scenario, there simply cannot
be a coherent joint probability distribution. At most certain subsets
of the relevant measurement outcomes can be jointly assigned coherent
probabilities.

As we have mentioned, however, non-fragmentation is central to much
of science, logic, and philosophy. This is so much the case that it
is rarely even mentioned as an explicit assumption. Accordingly, a
rejection of non-fragmentation comes, \emph{prima facie}, with considerable
cost. Before committing to such a route, many would want to have a
sensible fragmentalist alternative in hand. Indeed, there has been
some work to develop fragmentalist views in philosophy, including
in recent debates about consciousness and first-person facts (see,
e.g., Fine 2005, Lipman 2023, cf. Merlo 2016). Merlo (2016, p. 324),
for instance, writes (without endorsing the position): ``one could
take all points of view to be on a par vis-a-vis truth \emph{simpliciter}
by treating them as different \textquoteleft fragments\textquoteright{}
of an overall incoherent totality of facts''.

Fragmentalist views have been more readily adopted in the foundations
of physics so far. Within the Copenhagen family of interpretations,
Bohr's complementarity pushes us in this direction. The quantum logic
tradition (Wilce 2024) postulates that possible facts form an orthomodular
lattice, but not a Boolean algebra. Bub and Pitowsky (2010) similarly
point to non-Boolean event structures, emphasizing that while there
is one objective world, it does not support a classical, global fact
structure. With less emphasis on modifications to logic, the consistent
histories approach (Griffiths 2024) proposes that the principal distinction
of quantum mechanics lies in the availability of multiple mutually-incompatible
modes of description, so-called frameworks, such that no single globally
coherent totality of facts can obtain. Abramsky and Brandenburger
(2011) provide a formalization of a fragmentalist reading of contextuality
with sheaf theory.

\subsection{Rejecting one world}

If one wishes to retain measurement realism, non-relationalism, and
non-fragmentation (as well as locality, measurement independence,
and non-solipsism), one could take the final escape route from the
heptalemma, which is to reject the thesis of one world. Rejecting
this would be to deny that reality is exhausted by one objective world.
There must be more worlds, more ``seats'' for facts. We would have
to give up the assumption that ``reality'' and ``the world'' are
synonymous, since reality would consist of several distinct worlds,
which are in some sense parallel to each other but equally real. Each
world would be the total collection of facts that hold at that world.
Any such world can then be both coherent and non-relational. And if
there is more than one world, these could index different probability
measures, so that we no longer have the single joint probability distribution
of Bell's theorem, which gives rise to the no-go result.

A natural way to reject one world is to postulate \emph{subjective}
worlds, perhaps one for each observer. These could be worlds in which
genuinely subjective or first-personal facts reside. A subjective
world could be defined as what philosophers and logicians call a ``centered
world'', i.e., as an ordered pair consisting of an objective world
and a subject or agent around which it is conceptually ``centered''.
Or it could be understood in some primitive or basic sense, according
to which the notion of a subjective world is fundamental and the notion
of an objective world is merely emergent, arising from invariances
across subjective worlds. For example, any fact that is invariant
under changes in the observer could be deemed objective. This many-worlds
perspective has recently been discussed by one of us, though in relation
to consciousness rather than quantum mechanics (List 2023). Earlier
philosophical accounts that reject the assumption of one world, in
a variety of different ways, include Goodman's view that there is
a plurality of actual worlds (1978, 1984; cf. Declos 2019), Vacariu's
account of ``epistemologically different worlds'' (2005, 2016),
who also proposes its application to quantum mechanics, Honderich's
proposal that there are ``subjective physical worlds'' (2014), and
Gabriel's critique of the notion of a single world as a totality of
everything (2015).

The view in quantum mechanics that comes closest to an outright rejection
of one world is QBism. In fact, we will argue below that more explicitly
aligning with this picture is one of two generic directions in which
QBism might develop in the future.

\section{Where do Copenhagen, Everett, and QBism fit?}

Since some interpretations of quantum mechanics seem harder to categorize
in terms of our heptalemma than others, we now comment on three particularly
salient cases.

\subsection{Copenhagen}

Strictly speaking, there is no unique Copenhagen interpretation, but
only a family of overlapping views with a shared intellectual history
going back to work done in Copenhagen in the 1920s (Faye 2024). When
one speaks of \emph{the} Copenhagen interpretation, one could be thinking
of the views of any of a substantial list of physicists including
Bohr, Heisenberg, Born, Pauli, Peierls, Landau, Rosenfeld, von Weizsäcker,
and Peres. One could also have in mind a somewhat vague interpretational
gloss that is broadly inspired by these and other authors' ideas.
This makes ``the'' Copenhagen interpretation harder to pin down
in our heptalemma as well. Notwithstanding this diversity, it remains
a coherent enough family to robustly reject Bell's realism. Moreover,
any set of ideas that is recognizably Copenhagen will certainly wish
to preserve measurement independence and non-solipsism. Preservation
of locality is almost universal as well, although there are some exceptions,
notably Peres (1993).

Precisely how Copenhagen rejects Bell's realism does not appear to
be uniform across the family, but we can still narrow it down significantly.
First, and most straightforwardly, no Copenhagener seems to have entertained
the idea that there could be more than the one objective world. While
Pauli, for instance, takes issue with the notion of a ``detached
observer'', he does not go so far as to say that outcomes are personal
and don't obtain in an objective, public world.\footnote{The reader is encouraged to peruse a correspondence between Pauli
and Bohr, discussed by Fuchs (2017a, Section 3).} Second, it seems that Copenhagen does\linebreak{}
 not reject non-relationalism, either. While Bohr emphasized that
measurement quantities are only meaningful relative to the complete
experimental arrangement, this is best viewed either as a somewhat
operationalist attitude or as a form of relativism about description,
but not as a relational account of facts of the kind now endorsed
by Rovelli and others. Measurement outcomes, for Bohr, are always
distinctly permanent marks on photographic plates or instances of
irreversible amplification in large, classical measurement instruments
such that the results are objectively available for anyone to verify.

Thus, the way in which Copenhagen rejects Bell's realism appears to
be either by rejecting measurement realism entirely or by rejecting
non-fragmentation. Both readings are defendable and have been argued
for in the past. Almost all varieties of Copenhagen emphasize that
the quantum state is not an objective property of the system, and
is instead a mathematical tool for thinking about it. As this does
not appear to be a claim of subjectivity or relativism, this stance
can be read as a rejection of measurement realism through the denial
that there are well-defined probabilities associated to measurement
outcomes. This is often identified as the salient anti-realist feature
of the Copenhagen interpretation, see, e.g., Maudlin (2019). On the
other hand, the lesson that Bohr and others draw from the complementarity
principle is one in which a single picture does not exist that unites
outcomes that could be obtained in different experimental contexts
(Bohr 1963). This kind of reading seems to reject non-fragmentation.

\subsection{Everett}

The Everett interpretation is popularly known as ``the many-worlds
interpretation'' and was named ``the relative state formulation''
by its inventor. From the perspective of the heptalemma, however,
we suggest it should be viewed neither as a many-worlds interpretation
nor as a relational one. Rather, it arguably upholds both one world
and non-relationalism. The supposed ``many worlds'' of the Everett
interpretation are just branches of the universal wavefunction evolution,
and this straightforwardly fits into a single world, albeit one that
is richer than conventionally assumed (namely, branching-tree-like).
Chalmers (1996, p. 347) describes the Everett interpretation as a
\textquotedblleft one-big-world interpretation\textquotedblright :
\textquotedblleft There is only one world, but it has more in it than
we might have thought\textquotedblright{} (1996, p. 347). Similarly,
while speaking of a particular measurement outcome is only meaningful
relative to the branch in which it occurs, the ontology of this view
is that \emph{all} possible outcomes actually do occur. Although we
find ourselves in one branch, all branches are equally real. All of
these outcomes factually obtain, \emph{simpliciter}, in one coherent
world. The label ``multiverse'' arguably describes this idea reasonably
well, as long as it is understood that the multiverse in its entirety
is a single world in the philosophical sense, not many.

In the taxonomy of our heptalemma, we think the Everettians are best
understood as rejecting measurement realism. Since all possible outcomes
equally occur in the Everettian multiverse, albeit in different branches,
there isn't a fact of the matter as to which particular outcome actually
results from a given measurement. At most, Everettians could say that
\emph{all} of those outcomes are facts. But then there is no single
outcome for which one can have well-defined probabilities. Indeed,
probabilities are not straightforwardly part of Everettianism. On
this picture, Everettians reject measurement realism in the second
way discussed above. What is real according to the Everett interpretation
is the universal wavefunction and its unitary evolution. As in Quantum
Darwinism, measurement is an add-on, which is \textendash{} strictly
speaking \textendash{} not part of the physics except insofar as one
should address how measurement can be understood to be part of the
physical dynamics. Making contact with probabilities and with the
Born rule is a research project for Everettians and has been approached
from a variety of perspectives including decision theory (Deutsch
1999, Wallace 2012) and so-called self-locating probability (Saunders
1998, Sebens and Carroll 2018).

Could one also squeeze the Everett interpretation into different horns
of our heptalemma, by understanding it in a slightly different way?
One might say, for instance, that it gives up non-relationalism by
making facts relative to branches. But the Everett interpretation
is not normally understood as being committed to an ontology of dyadic
facts in the way RQM is. Alternatively, one might say that the Everett
interpretation postulates branch-specific facts, which are best viewed
as \emph{indexical} facts, akin to facts about where I am currently
located or what time it is now. Assuming I find myself in a particular
branch, it is an indexical fact that my measurement outcome is such
and such. One might then say that each branch corresponds to a different
total collection of indexical facts: a different ``indexical world''.
The Everett interpretation could thus be viewed as rejecting one world,
after all, by postulating many indexical worlds. However, we suspect
most Everettians would not want to ``reify'' those indexical worlds
or view them as primitives of their physical ontology. Indeed, as
Wallace (2012) has argued, the multiverse is an emergent phenomenon;
it is not ontologically fundamental.

\subsection{QBism}

QBism starts out by taking a radically anti-objectivist stance towards
quantum theory. At its most basic, it treats the quantum formalism
as an aid to an agent's decision-making. This sharply contrasts with
interpretations that suppose that quantum mechanics seeks to describe
the objective world. QBism asserts that quantum states and measurements
do not represent anything that transcends the user's unique situation,
but that they capture the user's capacity for action, physical embodiment,
and any beliefs the user holds. On this picture, the objects of quantum
mechanics are personal judgments that help the user to navigate the
consequences of his or her actions (where a consequence is understood
to be one's personal experience of a measurement outcome). Moreover,
QBists take themselves to interpret quantum theory \emph{normatively},
namely as providing checks on the consistency of an agent's beliefs,
similar to checking whether a set of credences form a coherent set
of probabilistic beliefs.

This much, however, is only part of the story. The greater part of
QBism's ambition, and where it remains a work in progress, is the
articulation of lessons about the nature of reality. QBism can thus
also be viewed as an ontological project. In particular, it asks,
what is it about reality that makes \emph{this} reasoning structure,
this particular mathematical addition to a physics-agnostic decision
theory, the right means for navigation. So, what is QBism's take on
our seven theses?

We noted above how QBism responds to accusations of solipsism. Only
a caricatured radical single-user version of QBism would be solipsistic.
Moreover, while it denies talk of hidden variables, QBism accepts
measurement independence insofar as it affirms that the quantum-mechanical
formalism can be used to help agents manage the consequences of their
\emph{freely chosen} actions. Locality, meanwhile, becomes almost
tautological because, in QBism, the quantum formalism never describes
physical influences between distant events; it is a tool for an agent
to manage her own possible future experiences, and these always occur
along her worldline, never at spacelike separations.\footnote{In this context, the equivalent of non-locality would have to be detected
\emph{within} the agent's mesh of beliefs in the form of a lack of
independence of her probability assignments for the outcomes of her
future actions on a distant system from her \emph{freely chosen} actions
on a nearby system.}

These affirmations require QBism to reject Bell's realism, which it
naturally does. Here, the heptalemma's refinement of Bell's realism
brings more clarity to QBism's conceptual location. The first thing
to highlight is that QBism strongly affirms non-relationalism, allowing
us to formalize the often-neglected distinction between QBism and
interpretations that admit relational facts, such as relational quantum
mechanics.\footnote{Pienaar (2021) and Barzegar and Oriti (2024) both describe QBism as
relational, but this appears to stem from lacking the distinctions
we draw in the heptalemma. As we have emphasized, a genuine rejection
of non-relationalism entails a reconstrual of the notion of a fact
as dyadic rather than monadic. RQM explicitly endorses this, but we
do not take QBism to do the same.} Fuchs, for instance, is clear that he rejects any kind of relationalism
when he writes:
\begin{quotation}
\noindent ``Experience is a far richer notion than a perspective
on some pre-existing thing. As well it is far more than a simple relation
among pre-existing things. Lived experience has an autonomy that neither
of these notions capture.'' (Fuchs 2023, p. 128)
\end{quotation}
But what about the other three theses that jointly entail Bell's realism?
Does QBism affirm measurement realism as we have stated it in the
heptalemma? While there is some subtlety, we think QBism is not only
consistent with measurement realism, but should be read as robustly
affirming it. There is no doubt that it regards measurement outcomes
as factual in the sense that outcomes are something that is the case.
Outcomes are intrinsically \emph{personal}, but they are not illusions,
relational, or approximate notions. The facts in question may be first-personal,
but they are facts no less.\footnote{Importantly, they are of the usual monadic form: such and such is
the case; they are not merely of the relational form: such and such
is the case relative to such and such.} Indeed, QBism looks to measurement for a grounding of the notion
of \emph{participatory} \emph{realism} (Fuchs 2017b). A recent paper
further foregrounds QBism's measurement realism, and situates it as
pursuing the diametric opposite of most approaches to the quantum
measurement problem, by deriving quantum dynamics from consistent
reasoning about measurement (DeBrota \emph{et al.} 2024). In other
words, rather than being somehow part of the dynamics, measurement
is the \emph{most} real thing according to QBism.

The subtlety comes from the second sense in which measurement realism
can be rejected, which corresponds to the denial that there are well-defined
probabilities for outcomes. QBism holds that there are well-defined
probabilities, but it treats them, like measurement outcomes, as personal
or subjective. QBists tend to view probabilities as degrees of belief
for the consequences of one's actions and even espouse a particularly
radical form of subjective Bayesianism that identifies probabilities
with betting dispositions. It is here that QBism might seem most anti-realist.

However, we suggest that such a purely epistemic or practical stance
on probabilities is not strictly \emph{mandated} by the rest of QBism's
commitments, at least according to our heptalemma. Our analysis shows
that there is no conflict, in principle, between a subjectivist interpretation
of probabilities and a realist one, in the weak sense of measurement
realism. The relevant probabilities can be both factual and subjective.

We think that the most straightforward version of QBism is one that
rejects one world while retaining measurement realism. Different observers'
measurements will then correspond to facts over which probabilities
are well defined; it just so happens that the totality of facts and
probability assignments can vary from observer to observer. Each observer
is associated with his or her own observer-specific world (in analogy
to the first-personally centered worlds discussed in List 2023). The
denial that reality is exhausted by one objective world would be consistent
with the above-mentioned philosophical critiques of one world (Goodman
1978, 1984; Vacariu 2005, 2016; Gabriel 2015). Mermin, especially,
seems to take this route:
\begin{quotation}
\noindent ``The fact is that \emph{my} science has a subject (me)
as well as an object (my world). \emph{Your} science has a subject
(you) as well as an object (your world). \emph{Alice}\textquoteright s
science has a subject (she) as well as an object (her world). I make
the same point three times to underline both the plurality of subjects,
and the plurality of worlds that each of us constructs on the basis
of our own individual experience.'' (Mermin 2019, p. 5)
\end{quotation}
Fuchs also seems to support a rejection of one world when he writes:
\begin{quotation}
\noindent ``{[}T{]}he lesson of quantum theory is that \emph{experience
happens}, and the refinement of the notion that comes from Wigner\textquoteright s
thought experiment is that there \emph{is a sense} in which those
experiences need not live in a single universe.'' (Fuchs 2023, p.
125)
\end{quotation}
Although both seem to be saying that the experiences of different
participants inhabit distinct worlds, our heptalemma allows us to
pinpoint more precisely what this view amounts to.

On the other hand, QBists also routinely speak of \emph{the} world
\textendash{} an unfinished, yet singular, world constantly coming
into being, where ``{[}e{]}ach quantum measurement creates something
new in the universe that is above and beyond the agent\textquoteright s
relation to the quantum system they are acting upon'' (Fuchs 2023,
p. 128). This intuition deprecates pre-existent subjective worlds
just as much as a pre-existent objective one. The idea is that reality
does not take place in any such containers, but is actively constituted
by parts that always maintain some autonomy, as in James's republican
banquet (James 1882). Indeed, QBism feels the pull of fragmentation.
For instance, Fuchs (2007) makes the analogy to Escher's paintings
of impossible objects. This reading locates the personalism of QBism
not in separate worlds, but in the sense that there are local parts
of the world which do not admit a global coherence. It seems, then,
that rather than rejecting one world, QBists could be saying that
the single world we live in is not a coherent totality.

If both of these are plausible readings, then, for now, there are
two distinct versions of QBism, in addition to the implausible solipsistic
version. The first version, which we may label ``Pluriverse QBism''\footnote{It is ironic that the word ``multiverse'' is already associated
with the Everett interpretation, as that interpretation arguably affirms
the one world thesis. In our treatment, if any interpretation deserved
the term ``many worlds'', it would be the version of QBism that
rejects one world. On the other hand, ``pluriverse'' is uniquely
appropriate for QBism due to its intellectual debts to William James.}, rejects one world and takes different agents to inhabit or experience
genuinely distinct subjective worlds that overlap in suitable ways.
The second version of QBism, which we may label ``Fragmentalist QBism'',
locates all observers in a single world, fundamentally fragmented
along suitable notions of personalist fault lines.

There is a clear formal difference between these two versions of QBism.
The formal machinery that would be needed to precisify each would
look very different. The pluriverse version would postulate many subjective
worlds and many subjective probability measures, where each of them
individually behaves in the standard way, i.e., each subjective world
is a complete and consistent collection of facts that hold at that
world, and each subjective probability measure is a coherent assignment
of probabilities to all elements of the relevant algebra of events.
By contrast, the fragmentalist version would not postulate any coherent
totalities at all. We would be dealing with a single world/reality
that is an incoherent totality and of which only fragments are coherent.
Moreover, we wouldn't have a globally coherent probability assignment
but only locally coherent subassignments. Pluriverse QBism might be
easier to develop in some ways, as it can probably leverage more from
existing philosophical literature. On the other hand, Fragmentalist
QBism might be relatively more at home among the variety of fragmentalist
constructions that have appeared in physics. Or, finally, it might
turn out that one or the other version of QBism is ultimately incompatible
with other convictions once considered closely enough.

\section{The heptalemma as a diagnostic criterion}

As noted, the seven theses are not jointly inconsistent by themselves,
but only together with the predictions of quantum mechanics. If physical
reality were classical \textendash{} i.e., in line with a pre-quantum-mechanical
worldview \textendash{} there would be no conflict between them; they
could be jointly satisfied. Had Einstein's theories of relativity
been the final word, in some sense, we would be in a position to accept
all of them. This observation suggests that the heptalemma may also
be put to another use: the seven theses give us a helpful diagnostic
criterion for determining whether a given scientific domain (as represented
by a particular scientific theory) should count as classical or not,
and if not, in which sense the given domain behaves non-classically.

For example, suppose we focus solely on a sufficiently macroscopic
(and non-quantum) subdomain of physics, or on a particular domain
within geology, or on a particular domain of the biomedical sciences.
Then it is quite plausible to think that we can consistently embrace
all seven theses (and we suspect that researchers in those fields
do so, at least implicitly): measurement outcomes correspond to real
facts, and there are well-defined probabilities over them; facts are
non-relational; the total collection of facts is coherent; there is
one objective world; everything is local; we have measurement independence;
and we can have as many observers as we like. The domain of quantum
mechanics stands out insofar as it is not like this. Whether quantum
mechanics is the only \textendash{} or the main \textendash{} example
of a non-classical domain in the sciences is a matter for debate.
Some philosophical accounts of consciousness, but by no means all,
suggest that the study of consciousness might be another non-classical
domain. Recall, in particular, the theories that postulate first-person
facts mentioned earlier.

Formally, the criterion for classicality is the following:

\medskip{}

\noindent\textbf{A criterion for classicality:} The domain of facts
represented by a particular scientific theory is classical if and
only if the theory in question is jointly consistent with all of measurement
realism, non-relationalism, non-fragmentation, one world, locality,
measurement independence, and non-solipsism.

\medskip{}

If a domain fails this test, this domain counts as non-classical,
and we can use the heptalemma to pinpoint the precise sense in which
it is non-classical, depending on the horn of the heptalemma that
we think best fits that domain. The ongoing debates about the interpretation
of quantum mechanics can be viewed as debates about this question.

\section{Concluding remarks}

We have presented a heptalemma for quantum mechanics: a seven-pronged
no-go result. One of its upshots is a novel taxonomy of interpretations
of quantum mechanics, where they are characterized in terms of which
horn of the heptalemma they take. Table 1 summarizes this taxonomy.
\begin{table}[h]
\begin{centering}
\caption{Interpretations of quantum mechanics and what they reject}
\par\end{centering}
\centering{}%
\begin{tabular}{|c|c|c|c|c|c|c|c|}
\hline 
{\tiny Interpretation} & \begin{cellvarwidth}[m]
\centering
{\tiny\vspace*{4bp}
}{\tiny\par}

{\tiny Measurement\vspace{-6bp}
}{\tiny\par}

{\tiny realism}
\end{cellvarwidth} & \begin{cellvarwidth}[m]
\centering
{\tiny\vspace*{4bp}
}{\tiny\par}

{\tiny Non-\vspace{-6bp}
}{\tiny\par}

{\tiny relationalism}
\end{cellvarwidth} & \begin{cellvarwidth}[m]
\centering
{\tiny\vspace*{4bp}
}{\tiny\par}

{\tiny Non-\vspace{-6bp}
}{\tiny\par}

{\tiny fragmentation}
\end{cellvarwidth} & \begin{cellvarwidth}[m]
\centering
{\tiny\vspace*{4bp}
}{\tiny\par}

{\tiny One\vspace{-6bp}
}{\tiny\par}

{\tiny world}
\end{cellvarwidth} & {\tiny Locality} & \begin{cellvarwidth}[m]
\centering
{\tiny\vspace*{4bp}
}{\tiny\par}

{\tiny Measurement\vspace{-6bp}
}{\tiny\par}

{\tiny independence}
\end{cellvarwidth} & \begin{cellvarwidth}[m]
\centering
{\tiny\vspace*{4bp}
}{\tiny\par}

{\tiny Non-\vspace{-6bp}
}{\tiny\par}

{\tiny solipsism}
\end{cellvarwidth}\tabularnewline
\hline 
\hline 
{\tiny de Broglie-Bohm} &  &  &  &  & \xmark &  & \tabularnewline
\hline 
{\tiny Collapse models} &  &  &  &  & \xmark &  & \tabularnewline
\hline 
{\tiny Wavefunction realism} &  &  &  &  & \xmark &  & \tabularnewline
\hline 
{\tiny Transactional} &  &  &  &  & \xmark &  & \tabularnewline
\hline 
{\tiny Indivisible stochastic} &  &  &  &  & \xmark &  & \tabularnewline
\hline 
{\tiny Superdeterminism} &  &  &  &  &  & \xmark & \tabularnewline
\hline 
{\tiny Cellular automaton} &  &  &  &  &  & \xmark & \tabularnewline
\hline 
{\tiny Statistical contextual} &  &  &  &  &  & \xmark & \tabularnewline
\hline 
{\tiny Retrocausal} &  &  &  &  &  & \xmark & \tabularnewline
\hline 
{\tiny Copenhagen} & \xmark &  &  &  &  &  & \tabularnewline
\hline 
{\tiny Everett} & \xmark &  &  &  &  &  & \tabularnewline
\hline 
{\tiny Quantum Darwinism} & \xmark &  &  &  &  &  & \tabularnewline
\hline 
{\tiny Relational} &  & \xmark &  &  &  &  & \tabularnewline
\hline 
{\tiny Pragmatic} &  & \xmark &  &  &  &  & \tabularnewline
\hline 
{\tiny Brukner} &  & \xmark &  &  &  &  & \tabularnewline
\hline 
{\tiny Quantum logic} &  &  & \xmark &  &  &  & \tabularnewline
\hline 
{\tiny Bub-Pitowsky} &  &  & \xmark &  &  &  & \tabularnewline
\hline 
{\tiny Consistent histories} &  &  & \xmark &  &  &  & \tabularnewline
\hline 
{\tiny Sheaf contextual} &  &  & \xmark &  &  &  & \tabularnewline
\hline 
{\tiny Fragmentalist QBism} &  &  & \xmark &  &  &  & \tabularnewline
\hline 
{\tiny Pluriverse QBism} &  &  &  & \xmark &  &  & \tabularnewline
\hline 
{\tiny Radical single user} &  &  &  &  &  &  & \xmark\tabularnewline
\hline 
\end{tabular}
\end{table}

In this table, we only indicate which thesis each interpretation primarily
rejects, leaving open its stand on the remaining six theses. There
may be some debate about whether some interpretations reject more
than one thesis, and there may not always be a clearcut answer, insofar
as some interpretations are best viewed as families of interpretations
that come in different variants. The logic of the heptalemma, however,
forces each interpretation to reject only one of the seven theses.
So, in principle, there could be a consistent variant of each interpretation
that rejects solely the indicated thesis and accepts all others. Finally,
like Bell's original theorem, the heptalemma could be refined further,
by splitting some of the seven theses into further conjuncts. The
result would be an even more fine-grained taxonomy. We hope, however,
to have struck a balance between informativeness and parsimony.

In the long run, of course, getting a clear lay of the land obtains
its full value not merely in telling us where it is possible to go,
but, hopefully, in helping us make a choice on which route to take.
Someday the challenges that quantum mechanics posed to the scientific
worldview might be taught as history, rather than a state of affairs
that has now persisted for over a century. Perhaps then a new challenge
will be just around the corner.

\section*{References}
\begin{lyxlist}{XXX}
\item [{Abramsky,}] S., and A. Brandenburger (2011). ``The sheaf-theoretic
structure of non-locality and contextuality.''\emph{ New Journal
of Physics} 13: 113036.\vspace{-6bp}
\item [{Barandes,}] J. A. (2025). ``Quantum Systems as Indivisible Stochastic
Processes.'' arXiv.org, arXiv:2507.21192.\vspace{-6bp}
\item [{Barzegar,}] A., and D. Oriti (2024). ``Epistemic\textendash Pragmatist
Interpretations of Quantum Mechanics: A Comparative Assessment.''
\emph{Foundations of Physics} 54: 66.\vspace{-6bp}
\item [{Bell,}] J. S. (1964). ``On the Einstein-Podolsky-Rosen Paradox.''
\emph{Physics Physique Fizika} 1: 195\textendash 200.\vspace{-6bp}
\item [{Bell,}] J. S. (1985). ``The theory of local beables.'' \emph{Dialectica}
39(2): 86\textendash 96.\vspace{-6bp}
\item [{Berkovitz,}] J. (2007). ``Action at a distance in quantum mechanics.''
In E. N. Zalta and U. Nodelman (eds.), \emph{The Stanford Encyclopedia
of Philosophy} (Spring 2007 Edition), \textless https://plato.stanford.edu/archives/spr2007/entries/qm-action-distance/\textgreater .\vspace{-6bp}
\item [{Bohr,}] N. (1963). \emph{Essays 1958-1962 on Atomic Physics and
Human Knowledge.} New York (John Wiley \& Sons).\vspace{-6bp}
\item [{Brukner,\,\v{C}.\,(2018).}] ``A\,no-go\,theorem\,for\,observer-independent\,facts.''\,\emph{Entropy}\,20(5):\,350.\vspace{-6bp}
\item [{Bub,}] J., and I. Pitowsky (2010). ``Two dogmas about quantum
mechanics.'' In S. Saunders, J. Barrett, A. Kent, and D. Wallace
(eds.), \emph{Many Worlds? Everett, Quantum Theory, and Reality},
Oxford (Oxford University Press), pp. 433\textendash 459.\vspace{-6bp}
\item [{Budroni,}] C., A. Cabello, O. Gühne, M. Kleinmann, and J.-Å. Larsson
(2022). ``Kochen\textendash \linebreak{}
Specker contextuality.'' \emph{Reviews of Modern Physics}\,94(4):
045007.\vspace{-6bp}
\item [{Calosi,}] C., and T. Riedel (2024). ``Relational Quantum Mechanics
at the Crossroads.'' \emph{Foundations of Physics} 54: 74.\vspace{-6bp}
\item [{Chalmers,}] D. J. (1996). \emph{The Conscious Mind: In Search of
a Fundamental Theory}. Oxford (Oxford University Press).\vspace{-6bp}
\item [{Cirel'son,}] B. S. (1980). ``Quantum generalizations of Bell's
inequality.'' \emph{Letters in Mathematical Physics} 4(2): 93\textendash 100.\vspace{-6bp}
\item [{Clauser,}] J. F., M. A. Horne, A. Shimony, and R. A. Holt (1969).
``Proposed experiment to test local hidden\nobreakdash-variable
theories.'' \emph{Physical Review Letters} 23: 880\textendash 884.\vspace{-6bp}
\item [{Colbeck,}] R., and R. Renner (2011). ``No extension of quantum
theory can have improved predictive power.'' \emph{Nature Communications}
2: 411.\vspace{-6bp}
\item [{Cramer,}] J. G. (1986). ``The transactional interpretation of
quantum mechanics.'' \emph{Reviews of Modern Physics} 58(3): 647\textendash 687.\vspace{-6bp}
\item [{DeBrota,}] J. B., C. A. Fuchs, and R. Schack (2024). ``QBism\textquoteright s
account of quantum dynamics and decoherence.'' \emph{Physical Review
A} 110(5): 052205.\vspace{-6bp}
\item [{Declos,}] A. (2019). ``Goodman\textquoteright s Many Worlds.''
\emph{Journal for the History of Analytical Philosophy} 7: 6.\vspace{-6bp}
\item [{Deutsch,}] D. (1999). ``Quantum theory of probability and decisions.''\emph{
Proceedings of the Royal Society A} 455: 3129\textendash 3137.\vspace{-6bp}
\item [{Di\ Biagio,}] A, and C. Rovelli (2021). ``Stable Facts, Relative
Facts.'' \emph{Foundations of Physics} 51: 30.\vspace{-6bp}
\item [{Einstein,}] A. (1948). ``Quanten-Mechanik und Wirklichkeit.''
\emph{Dialectica} 2(3\textendash 4): 320\textendash 324.\vspace{-6bp}
\item [{Einstein,}] A., B. Podolsky, and N. Rosen (1935). ``Can quantum-mechanical
description of physical reality be considered complete?'' \emph{Physical
Review} 47(10): 777\textendash 780.\vspace{-6bp}
\item [{Faye,}] J. (2024). ``Copenhagen Interpretation of Quantum Mechanics.''
In E. N. Zalta and U. Nodelman (eds.), \emph{The Stanford Encyclopedia
of Philosophy} (Summer 2024 Edition), \textless https://plato.stanford.edu/archives/sum2024/entries/qm-copenhagen/\textgreater .\vspace{-6bp}
\item [{Fine,}] A. (1982). ``Joint distributions, quantum correlations,
and commuting observables.'' \emph{Journal of Mathematical Physics}
23(7): 1306\textendash 1310.\vspace{-6bp}
\item [{Fine,}] K. (2005). ``Tense and Reality.'' In \emph{Modality and
Tense: Philosophical Papers}, Oxford: Oxford University Press, pp.
261\textendash 320.\vspace{-6bp}
\item [{Friederich,}] S., and P. W. Evans (2023). ``Retrocausality in
Quantum Mechanics.'' \linebreak{}
In E. N. Zalta and U. Nodelman (eds.), \emph{The Stanford }\linebreak{}
\emph{Encyclopedia of Philosophy} (Winter 2023 Edition), \linebreak{}
\textless https://plato.stanford.edu/archives/win2023/entries/qm-retrocausality/\textgreater .\vspace{-6bp}
\item [{Fuchs,}] C. A. (2007). ``Delirium Quantum: Or, Where I Will Take
Quantum Mechanics If It Will Let Me.'' In G. Adenier, C. A. Fuchs,
and A. Yu. Khrennikov (eds.), \emph{Foundations of Probability and
Physics \textemdash{} 4, AIP Conference Proceedings Vol. 889}, Melville,
NY (American Institute of Physics), pp. 438\textendash 462.\vspace{-6bp}
\item [{Fuchs,}] C. A. (2017a). ``Notwithstanding Bohr, the Reasons for
QBism.'' \emph{Mind and Matter} 15(2): 245\textendash 300.\vspace{-6bp}
\item [{Fuchs,}] C. A. (2017b). ``On Participatory Realism.'' In I. T.
Durham and D. Rickles (eds.), \emph{Information and Interaction: Eddington,
Wheeler, and the Limits of Knowledge}, Dordrecht (Springer), pp. 113\textendash 134.\vspace{-6bp}
\item [{Fuchs,}] C. A. (2023). ``QBism, Where Next?'' In P. Berghofer
and H. A. Wiltsche (eds.), \emph{Phenomenology and QBism: New Approaches
to Quantum Mechanics}, New York (Routledge), pp. 78\textendash 143.\vspace{-6bp}
\item [{Fuchs,}] C. A., and R. Schack (2013). \textquotedblleft Quantum-Bayesian
coherence.\textquotedblright{} \emph{Reviews of Modern Physics} 85(4):
1693\textendash 1715.\vspace{-6bp}
\item [{Fuchs,}] C. A., and B. C. Stacey (2018). \textquotedblleft QBism:
Quantum Theory as a Hero\textquoteright s Handbook.\textquotedblright{}
In E. M. Rasel, W. P. Schleich, and S. Wölk (eds.), \emph{Proceedings
of the International School of Physics \textquotedblleft Enrico Fermi\textquotedblright{}
Course 197 \textendash{} Foundations of Quantum Physics}, Società
Italiana di Fisica, Bologna, Amsterdam (IOS Press), pp. 133\textendash 202.\vspace{-6bp}
\item [{Gabriel,}] M. (2015). \emph{Why the World Does Not Exist}. Cambridge
(Polity).\vspace{-6bp}
\item [{Ghirardi,}] G., A. Rimini, and T. Weber (1986). ``Unified dynamics
for microscopic and macroscopic systems.'' \emph{Physical Review
D} 34(2): 470\textendash 491.\vspace{-6bp}
\item [{Goldstein,}] S. (2025). ``Bohmian Mechanics.'' In E. N. Zalta
and U. Nodelman (eds.), \emph{The Stanford Encyclopedia of Philosophy}
(Fall 2025 Edition), \linebreak{}
\textless https://plato.stanford.edu/archives/fall2025/entries/qm-bohm/\textgreater .\vspace{-6bp}
\item [{Goodman,}] N. (1978). \emph{Ways of Worldmaking}. Indianapolis
(Hackett).\vspace{-6bp}
\item [{Goodman,}] N. (1984). \emph{Of Mind and Other Matters}. Cambridge,
MA (Harvard University Press).\vspace{-6bp}
\item [{Greenberger,}] D. M., M. A. Horne, and A. Zeilinger (1989). ``Going
beyond Bell's Theorem.'' In M. Kafatos (ed.), \emph{Bell's Theorem,
Quantum Theory and Conceptions of the Universe}, Dordrecht (Kluwer),
pp. 69\textendash 72.\vspace{-6bp}
\item [{Griffiths,}] R. B. (2024). \textquotedblleft The Consistent Histories
Approach to Quantum Mechanics.\textquotedblright{} In E. N. Zalta
and U. Nodelman (eds.), \emph{The Stanford Encyclopedia of Philosophy}
(Summer 2024 Edition), \textless https://plato.stanford.edu/archives/sum2024/entries/\linebreak{}
qm-consistent-histories/\textgreater .\vspace{-6bp}
\item [{Hall,}] N. (2004). \textquotedblleft Two Concepts of Causation.\textquotedblright{}
In J. Collins, N. Hall, and L. A. Paul (eds.), \emph{Causation and
Counterfactuals}, Cambridge, MA (MIT Press), pp. 225\textendash 276.\vspace{-6bp}
\item [{Hardy,}] L. (1992). ``Quantum mechanics, local realistic theories,
and Lorentz-invariant realistic theories.'' \emph{Physical Review
Letters} 68(20): 2981\textendash 2984.\vspace{-6bp}
\item [{Healey,}] R. (2012). ``Quantum theory: a pragmatist approach.''
\emph{British Journal for the Philosophy of Science} 63(4): 729\textendash 771.\vspace{-6bp}
\item [{Honderich,}] T. (2014). \emph{Actual Consciousness}. Oxford (Oxford
University Press).\vspace{-6bp}
\item [{Hossenfelder,}] S., and T. Palmer (2020). \textquotedblleft Rethinking
Superdeterminism.\textquotedblright{} \emph{Frontiers in Physics}
8: 139. \vspace{-6bp}
\item [{Howard,}] D. (1985). ``Einstein on Locality and Separability.''
\emph{Studies in History and Philosophy of Science Part A }16(3):
171\textendash 201.\vspace{-6bp}
\item [{James,}] W. (1882). ``On some Hegelisms.'' \emph{Mind} 7(26):
186\textendash 208.\vspace{-6bp}
\item [{Jarrett,}] J. (1984). ``On the Physical Significance of the Locality
Conditions in the Bell Arguments.'' \emph{Noûs} 18(4): 569\textendash 589.\vspace{-6bp}
\item [{Kochen,}] S., and E. P. Specker (1967). ``The Problem of Hidden
Variables in Quantum Mechanics.'' \emph{Journal of Mathematics and
Mechanics} 17(1): 59\textendash 87.\vspace{-6bp}
\item [{Kupczynski,}] M. (2025). ``Statistical contextual explanation
of quantum paradoxes.'' \emph{Frontiers in Quantum Science and Technology}
4: 1569496.\vspace{-6bp}
\item [{Leggett,}] A. J. (2003). ``Nonlocal Hidden-Variable Theories and
Quantum Mechanics: An Incompatibility Theorem.'' \emph{Foundations
of Physics} 33: 1469\textendash 1493.\vspace{-6bp}
\item [{Leifer,}] M. S. (2014). ``Is the quantum state real? An extended
review of \textgreek{\textpsi}\nobreakdash-ontology theorems.''\,\emph{Quanta}
3: 67\textendash 155.\vspace{-6bp}
\item [{Lipman,}] M. (2023). ``Subjective Facts about Consciousness.''
\emph{Ergo} 10: 530\textendash 553.\vspace{-6bp}
\item [{List,}] C. (2023). ``The many-worlds theory of consciousness.''
\emph{Noûs} 57(2): 316\textendash 340.\vspace{-6bp}
\item [{List,}] C. (2025). ``A quadrilemma for theories of consciousness.''
\emph{The Philosophical Quarterly} 75(3): 1026\textendash 1048.\vspace{-6bp}
\item [{Maudlin,}] T. (2019). \emph{Philosophy of Physics: Quantum Theory.}
Princeton, NJ (Princeton University Press).\vspace{-6bp}
\item [{Merlo,}] G. (2016). \textquotedblleft Subjectivism and the mental.\textquotedblright{}
\emph{Dialectica} 70(3): 311\textendash 342.\vspace{-6bp}
\item [{Mermin,}] N. D. (2019). ``Making better sense of quantum mechanics.''
\emph{Reports on Progress in Physics} 82: 012002.\vspace{-6bp}
\item [{Myrvold,}] W., M. Genovese, and A. Shimony (2024). ``Bell\textquoteright s
Theorem.'' In E. N. Zalta and U. Nodelman (eds.), \emph{The Stanford
Encyclopedia of Philosophy} (Spring 2024 Edition), \textless https://plato.stanford.edu/archives/spr2024/entries/bell-theorem/\textgreater .\vspace{-6bp}
\item [{Newton,}] I. (1692/3). Letter to Richard Bentley. The Newton Project.
\linebreak{}
\textless https://www.newtonproject.ox.ac.uk/catalogue/record/THEM00258\textgreater .\vspace{-6bp}
\item [{Ney,}] A. (2021). \emph{The World in the Wavefunction: A Metaphysics
for Quantum Physics.} Oxford (Oxford University Press).\vspace{-6bp}
\item [{Nikoli\'{c},}] H. (2012). ``Solipsistic hidden variables.'' \emph{International
Journal of Quantum Information} 10(8): 1241016.\vspace{-6bp}
\item [{Norsen,}] T. (2014). ``Quantum Solipsism and Non-Locality.''
\emph{International Journal of Quantum Foundations}, \textless https://ijqf.org/archives/1548/\textgreater .\vspace{-6bp}
\item [{Pearl,}] J. (2000). \emph{Causality: Models, Reasoning, and Inference}.\emph{
}Cambridge (Cambridge University Press).\vspace{-6bp}
\item [{Peres,}] A. (1993). \emph{Quantum Theory: Concepts and Methods.}
Dordrecht (Kluwer Academic Publishers).\vspace{-6bp}
\item [{Pienaar,}] J. (2021). ``QBism and Relational Quantum Mechanics
compared.'' \emph{Foundations of Physics} 51: 96.\vspace{-6bp}
\item [{Pusey,}] M. F., J. Barrett, and T. Rudolph (2012). ``On the reality
of the quantum state.'' \emph{Nature Physics} 8(6): 475\textendash 478.\vspace{-6bp}
\item [{Price,}] H., and K. Wharton (2015). \textquotedblleft Disentangling
the quantum world.\textquotedblright{} \emph{Entropy} 17(11): 7752\textendash 7767.\vspace{-6bp}
\item [{Rovelli,}] C. (1996). ``Relational quantum mechanics.'' \emph{International
Journal of Theoretical Physics} 35(8): 1637\textendash 1678.\vspace{-6bp}
\item [{Rovelli,}] C. (2025). ``Relational Quantum Mechanics.'' In E.
N. Zalta and U. Nodelman (eds.), \emph{The Stanford Encyclopedia of
Philosophy} (Spring 2025 Edition), \linebreak{}
\textless https://plato.stanford.edu/archives/spr2025/entries/qm-relational/\textgreater .\vspace{-6bp}
\item [{Saunders,}] S.\,(1998). \textquotedblleft Time,\,quantum\,mechanics,\,and\,probability.\textquotedblright{}
\emph{Synthese}\,114:\,373\textendash 404.\vspace{-6bp}
\item [{Schmid,}] D., Y. Y\={\i}ng, and M. S. Leifer (2023). ``A review
and analysis of six extended Wigner\textquoteright s friend arguments.''
arXiv.org, arXiv:2308.16220.\vspace{-6bp}
\item [{Sebens,}] C. T., and S. M. Carroll (2018). ``Self-locating uncertainty
and the origin of probability in Everettian quantum mechanics.''
\emph{British Journal for the Philosophy of Science} 69(1): 25\textendash 74.\vspace{-6bp}
\item [{Stacey,}] B. C. (2018). ``Misreading EPR: Variations on an Incorrect
Theme.'' arXiv.org, arXiv:1809.01751.\vspace{-6bp}
\item [{Stacey,}] B. C. (2021). ``Is Relational Quantum Mechanics about
Facts? If So, Whose? A Reply to Di Biagio and Rovelli\textquoteright s
Comment on Brukner and Pienaar.'' arXiv.org, arXiv:2112.07830.\vspace{-6bp}
\item [{\textquoteright t\ Hooft,}] G. (2016). \emph{The Cellular Automaton
Interpretation of Quantum Mechanics.} Vol. 185. Fundamental Theories
of Physics. Cham (Springer).\vspace{-6bp}
\item [{Vacariu,}] G. (2005). ``Mind, Brain, and Epistemologically Different
Worlds.'' \emph{Synthese} 147(3): 515\textendash 548.\vspace{-6bp}
\item [{Vacariu,}] G. (2016). \emph{Illusions of Human Thinking: On Concepts
of Mind, Reality, and Universe in Psychology, Neuroscience, and Physics}.
Wiesbaden (Springer).\vspace{-6bp}
\item [{Vaidman,}] L. (2021). ``Many-Worlds Interpretation of Quantum
Mechanics.'' In E. N. Zalta (ed.)\emph{, The Stanford Encyclopedia
of Philosophy} (Fall 2021 Edition), \textless https://plato.stanford.edu/archives/fall2021/entries/qm-manyworlds/\textgreater .\vspace{-6bp}
\item [{Wallace,}] D. (2012). \emph{The Emergent Multiverse: Quantum Theory
according to the Everett Interpretation.} Oxford (Oxford University
Press).\vspace{-6bp}
\item [{Wigner,}] E. P. (1961). ``Remarks on the Mind-Body Question.''
In I. J. Good (ed.), \emph{The Scientist Speculates: An Anthology
of Partly-Baked Ideas}, London (Heinemann), pp. 284\textendash 302.\vspace{-6bp}
\item [{Wilce,}] A. (2024). ``Quantum Logic and Probability Theory.''
In E. N. Zalta and U. Nodelman (eds.), \emph{The Stanford Encyclopedia
of Philosophy} (Summer 2024 Edition), \textless https://plato.stanford.edu/archives/sum2024/entries/qt-quantlog/\textgreater .\vspace{-6bp}
\item [{Wittgenstein,}] L. (1922) \emph{Tractatus Logico-Philosophicus}.
London (Kegan Paul).\vspace{-6bp}
\item [{Zurek,}] W. H. (2009). ``Quantum Darwinism.'' \emph{Nature Physics}
5: 181\textendash 188.\vspace{-6bp}
\item [{Zwirn,}] H. (2000). \emph{Les limites de la connaissance.} Paris
(Odile Jacob).\vspace{-6bp}
\end{lyxlist}

\section*{Appendix: Bell's theorem}

In this appendix we present a pedagogical derivation of Bell's theorem.
The following does not trace one particular source, though many have
appeared. One can refer to Myrvold \emph{et al.} (2024) for further
details.

The Bell scenario concerns two spacelike separated agents Alice and
Bob, each equipped with a measurement device. A source produces two
signals, sending one to Alice's device and one to Bob's. Each agent
chooses a measurement setting $a$ and $b$ and subsequently registers
an outcome $A$ and $B$.

Bell's theorem establishes that quantum theory is not compatible with
the conjunction of the following theses:\smallskip{}

\noindent\textbf{Realism:} There exists a single joint probability
space such that all measurement outcomes $A$ and $B$ are probabilistically
determined by the values of the measurement settings $a,b$ and a
``hidden variable'' $\lambda$.\smallskip{}

\noindent\textbf{Locality:} Measurements performed in one place do
not instantaneously bring about effects in a distant place. So, the
outcome $B$ does not depend on the setting $a$ or the outcome $A$;
likewise, the outcome $A$ does not depend on the setting $b$ or
the outcome $B$.\smallskip{}

\noindent\textbf{Measurement independence:} Measurement choices are
statistically independent of each other and of the system being measured.
This ensures that the experimenters' choices are not pre-determined
by the physical scenario.\smallskip{}

For simplicity we assume that the probability space is discrete, so
as to avoid the potentially subtle machinery of measure theory, but
this comes with no conceptual loss.

Realism ensures that there are objective facts about the five variables
$\lambda$, $A$, $B$, $a$, $b$ and their probabilistic dependencies.
The dependency structure entailed by locality and measurement independence
then results in a factorization of the joint probability for $A$
and $B$ given the measurement choices $a,b$. Let us spell this out.
First, the objective reality of $\lambda$ ensures this joint probability
can be expressed as a single marginalization over the possible hidden
variable values, 
\begin{equation}
P(A,B|a,b)=\sum\nolimits_{\lambda}P(A,B,\lambda|a,b)=\sum\nolimits_{\lambda}P(\lambda|a,b)P(A,B|a,b,\lambda),
\end{equation}
where the second equality follows from the product rule. Locality
implies the independence of $A$ and $B$ from settings or outcomes
on the other side 
\begin{equation}
P(A|a,b,B,\lambda)=P(A|a,\lambda)\quad\text{and}\quad P(B|a,b,A,\lambda)=P(B|b,\lambda),
\end{equation}
so that, after another use of the product rule, we see that 
\begin{equation}
P(A,B|a,b,\lambda)=\begin{cases}
P(A|a,b,\lambda)P(B|a,b,A,\lambda)\\
P(A|a,b,B,\lambda)P(B|a,b,\lambda)
\end{cases}\!\!\!=P(A|a,\lambda)P(B|b,\lambda).
\end{equation}
Measurement independence establishes that the probability of $\lambda$
cannot depend on the measurement settings $a,b$ the experimenters
choose, 
\begin{equation}
P(\lambda|a,b)=P(\lambda).
\end{equation}
These observations together imply the factorization 
\begin{equation}
P(A,B|a,b)=\sum\nolimits_{\lambda}P(\lambda)P(A|a,\lambda)P(B|b,\lambda).\label{factorized}
\end{equation}
The factorized form \eqref{factorized} is then used to derive so-called
Bell inequalities, which may be understood as bounds on the strength
of measurement outcome correlations compatible with Bell's theses.
Here we present the well-known CHSH inequality (Clauser\emph{\,et\,al.\,}1969).

Suppose Alice and Bob each choose between two measurement settings,
which (in a slight abuse of notation\footnote{Here, $a$ and $a'$ are the possible values for the variable $a$,
and $b$ and $b'$ are those for the variable $b$.}) we denote $a,a'$ on Alice's side and $b,b'$ on Bob's side, and
that each measurement has only two outcomes $A,B\in\{\pm1\}$. For
$x\in\{a,a'\}$ and $y\in\{b,b'\}$, define the correlation, the expected
product of the outcomes, 
\begin{equation}
E(x,y):=\sum_{A,B=\pm1}A\;B\;P(A,B|x,y),\label{correlation}
\end{equation}
and the conditional expectations 
\begin{equation}
\bar{A}(x,\lambda):=\sum_{A=\pm1}A\;P(A|x,\lambda),\quad\bar{B}(y,\lambda):=\sum_{B=\pm1}B\;P(B|y,\lambda).
\end{equation}
Substituting \eqref{factorized} into \eqref{correlation}, we see
that Bell's assumptions imply that the correlation is the averaged
product of the expectations 
\begin{equation}
E(x,y)=\sum\nolimits_{\lambda}P(\lambda)\bar{A}(x,\lambda)\bar{B}(y,\lambda).
\end{equation}

For simplicity, we first assume that $\lambda$ and the measurement
setting deterministically determine the measurement outcomes, that
is, that the conditional probabilities take only the values $P(A|x,\lambda)\in\{0,1\}$
and $P(B|y,\lambda)\in\{0,1\}$, so the conditional expectations are
equal to the outcome values and 
\begin{equation}
A(a,\lambda),A(a',\lambda),B(b,\lambda),B(b',\lambda)\in\{\pm1\},
\end{equation}
so that the correlation function in this case is 
\begin{equation}
E(x,y)=\sum\nolimits_{\lambda}P(\lambda)A(x,\lambda)B(y,\lambda).
\end{equation}

Now we define the CHSH combination, which relates the correlations
associated with four different experiments, 
\begin{equation}
S:=E(a,b)+E(a,b')+E(a',b)-E(a',b').
\end{equation}
For a fixed $\lambda$ define 
\begin{equation}
S(\lambda):=A(a,\lambda)B(b,\lambda)+A(a,\lambda)B(b',\lambda)+A(a',\lambda)B(b,\lambda)-A(a',\lambda)B(b',\lambda).
\end{equation}
Factor $S(\lambda)$ as 
\begin{equation}
S(\lambda)=A(a,\lambda)\bigl[B(b,\lambda)+B(b',\lambda)\bigr]+A(a',\lambda)\bigl[B(b,\lambda)-B(b',\lambda)\bigr].\label{CHSHfactored}
\end{equation}
Since $B(b,\lambda),B(b',\lambda)\in\{\pm1\}$, there are two cases:
\begin{enumerate}
\item If $B(b,\lambda)=B(b',\lambda)$, then $B(b,\lambda)+B(b',\lambda)=\pm2$
and $B(b,\lambda)-B(b',\lambda)=0$, so $S(\lambda)=\pm2A(a,\lambda)$.
\item If $B(b,\lambda)=-B(b',\lambda)$, then $B(b,\lambda)+B(b',\lambda)=0$
and $B(b,\lambda)-B(b',\lambda)=\pm2$, so $S(\lambda)=\pm2A(a',\lambda)$.
\end{enumerate}
In all cases $|S(\lambda)|=2$. Averaging over $\lambda$ with $P(\lambda)$
gives 
\begin{equation}
|S|=\Bigl|\sum\nolimits_{\lambda}P(\lambda)\,S(\lambda)\Bigr|\le\sum\nolimits_{\lambda}P(\lambda)|S(\lambda)|=2.
\end{equation}
Therefore, for any \emph{deterministic} theory satisfying the Bell
theses, 
\begin{equation}
|E(a,b)+E(a,b')+E(a',b)-E(a',b')|\le2,\label{eq:CHSH}
\end{equation}
which is the CHSH inequality.

This proof straightforwardly extends to the stochastic case where
we place no restrictions on the conditional probabilities. Replace
the values in equation \eqref{CHSHfactored} with their conditional
means $\bar{A}(x,\lambda)$ and $\bar{B}(y,\lambda)$. Then $|\bar{A}(y,\lambda)|\le1$
implies 
\begin{equation}
|S(\lambda)|\le|\bar{B}(b,\lambda)+\bar{B}(b',\lambda)|+|\bar{B}(b,\lambda)-\bar{B}(b',\lambda)|.
\end{equation}
Now for two real numbers $x,y$ with $|x|\leq1,|y|\leq1$, one has
$|x+y|+|x-y|\leq2$.\footnote{If $x$ and $y$ have the same sign, then $|x+y|=|x|+|y|$ and $|x-y|=||x|-|y||$
and it follows that $|x+y|+|x-y|=2\max(|x|,|y|)\leq2$. If they have
opposite signs, the roles reverse, but the bound is the same.} Applying this with $x=\bar{B}(b,\lambda)$ and $y=\bar{B}(b',\lambda)$
yields $|S(\lambda)|\leq2$ as before.

So, subject to the Bell assumptions, measurement outcome correlations
may not exceed the CHSH inequality bound of $2$. However, quantum
mechanics predicts a violation of this bound up to a maximal value
of $2\sqrt{2}$, as we now show.

Suppose Alice and Bob each receive a two-level quantum system and
each chooses between two possible quantum measurements with outcome
values of $\pm1$. This setting can be realized as measurements $\hat{a}=\vec{a}\cdot\vec{\sigma},\;\hat{a}'=\vec{a}'\cdot\vec{\sigma},\;\hat{b}=\vec{b}\cdot\vec{\sigma},\;\hat{b}'=\vec{b}'\cdot\vec{\sigma}$,
where $\vec{\sigma}=({\hat{{\sigma}}}_{x},{\hat{{\sigma}}}_{y},{\hat{{\sigma}}}_{z})$
is the vector of Pauli matrices and $\vec{a}$, $\vec{a}'$, $\vec{b}$,
$\vec{b}'$ are unit vectors in $\mathbb{R}^{3}$, so that Alice's
and Bob's measurement choices can each be considered choices between
two directions in space along which to measure spin. Given a pure
quantum state $\ket{\psi}$ on the bipartite system, the correlation
of measurement outcomes, say with measurement settings $a$ and $b$,
is $E(a,b)=\bra{\psi}\hat{a}\otimes\hat{b}\ket{\psi}$. Consider the
two-qubit singlet state 
\[
\ket{\psi^{-}}:=\frac{1}{\sqrt{2}}(\ket{01}-\ket{10}).
\]
For spin measurements, the singlet correlations are 
\[
\bra{\psi^{-}}(\vec{u}\cdot\vec{\sigma})\otimes(\vec{v}\cdot\vec{\sigma})\ket{\psi^{-}}=-\,\vec{u}\cdot\vec{v}.
\]
Thus, for this state, 
\[
S=-\big(\vec{a}\!\cdot\!\vec{b}+\vec{a}\!\cdot\!\vec{b}'+\vec{a}'\!\cdot\!\vec{b}-\vec{a}'\!\cdot\!\vec{b}'\big).
\]
Choosing the four unit vectors to be 
\[
\vec{a}=(1,0),\quad\vec{a}'=(0,1),\quad\vec{b}=\tfrac{1}{\sqrt{2}}(1,1),\quad\vec{b}'=\tfrac{1}{\sqrt{2}}(1,-1),
\]
we have 
\[
\vec{a}\!\cdot\!\vec{b}=\vec{a}\!\cdot\!\vec{b}'=\vec{a}'\!\cdot\!\vec{b}=\tfrac{1}{\sqrt{2}},\qquad\vec{a}'\!\cdot\!\vec{b}'=-\tfrac{1}{\sqrt{2}},
\]
so 
\[
S=-\Big(\tfrac{1}{\sqrt{2}}+\tfrac{1}{\sqrt{2}}+\tfrac{1}{\sqrt{2}}-(-\tfrac{1}{\sqrt{2}})\Big)=-\,\tfrac{4}{\sqrt{2}}=-\,2\sqrt{2}.
\]
Therefore $|S|=2\sqrt{2}$, which exceeds the bound predicted by Bell's
assumptions. The fact that this violation is maximal is known as Tsirelson's
bound (Cirel'son 1980) and follows from operator norm arguments that
we omit here.

To summarize, the Bell assumptions predict that measurement outcome
correlations cannot exceed the CHSH inequality bound of 2. Quantum
mechanics predicts a violation of this bound. Indeed, violations of
Bell inequalities have been conclusively demonstrated experimentally
as well. Consequently, at least one of Bell's assumptions must be
denied.
\end{document}